\documentclass[runningheads]{llncs}

\usepackage{amsthm,amsmath}
\usepackage{amsfonts}
\usepackage{graphicx}
\usepackage[utf8]{inputenc}
\usepackage{numprint}
\usepackage{todonotes}
\usepackage{hyperref} 
\usepackage{multirow} 
\usepackage[tableposition=top]{caption}

\npthousandsep{,}\npthousandthpartsep{}\npdecimalsign{.}

\begin{document}

\title{Identifying stress responsive genes using overlapping communities in co-expression networks}

\titlerunning{Stress responsive genes in co-expression networks}

\author{Camila Riccio \and Jorge Finke \and Camilo Rocha}
\authorrunning{C. Riccio et al.}

\institute{Pontificia Universidad Javeriana, Cali, Colombia
  \email{\{camila.riccio,jfinke,camilo.rocha\}@javerianacali.edu.co}
}  
\maketitle

\begin{abstract}

\textbf{Background} This paper proposes a workflow to identify
genes that respond to specific treatments in plants.  The workflow
takes as input the RNA sequencing read counts and phenotypical data of
different genotypes, measured under control and treatment conditions.
It outputs a reduced group of genes marked as relevant for treatment
response. Technically, the proposed approach is both a generalization
and an extension of WGCNA.  It aims to identify specific modules of
overlapping communities underlying the co-expression network of genes.
Module detection is achieved by using Hierarchical Link Clustering.
The overlapping nature of the systems' regulatory domains that
generate co-expression can be identified by such modules.  LASSO
regression is employed to analyze phenotypic responses of modules to
treatment.

\textbf{Results} The workflow is applied to rice (\textit{Oryza
  sativa}), a major food source known to be highly sensitive to salt
stress.  The workflow identifies 19 rice genes that seem relevant in
the response to salt stress.  They are distributed across 6 modules: 3
modules, each grouping together 3 genes, are associated to shoot K
content; 2 modules of 3 genes are associated to shoot biomass; and 1
module of 4 genes is associated to root biomass. These genes represent
target genes for the improvement of salinity tolerance in rice.
  
\textbf{Conclusions} A more effective framework to reduce the
search-space for target genes that respond to a specific treatment is
introduced.  It facilitates experimental validation by restraining
efforts to a smaller subset of genes of high potential relevance.

            
\end{abstract}

\section*{Introduction}
\label{sec.intro}

Stresses are key factors that influence plant development, often
associated to extensive losses in agricultural
production~\cite{mesterhazy2020losses,shrivastava2015soil}.  Soil
salinity is one of the most devastating abiotic stresses. According
to~\cite{shrivastava2015soil}, soil salinity contributes to a
significant reduction in areas of cultivable land and crop
quality. The study estimates that 20\% of the total cultivated land
worldwide and 33\% of the total irrigated agricultural land is
affected by high salinity.  By the end of 2050, areas of high salinity
are expected to reach 50\% of the cultivated
land~\cite{shrivastava2015soil}.

Salinity tolerance and susceptibility are the result of elaborated
interactions between morphological, physiological, and biochemical
processes. They are regulated by multiple genes in various parts of
the plant genome~\cite{reddy2017salt}. Consequently, identifying
groups of responsive genes is an important step for improving crop
varieties in terms of salinity tolerance.  This paper proposes a
workflow to identify stress responsive genes associated with a complex
quantitative trait.

To discover the genes associated with a phenotypic response to
treatment, the workflow takes as input the gene expression profiles of
the target organism. Specifically, it takes the RNA sequencing read
counts (measured under control and treatment conditions) of at least
two biological replicates per genotype. It also receives phenotypic
data in the form of observable traits, measured for each genotype
under the two conditions. The output of the workflow is a set of genes
that are characterized as potentially relevant to treatment.

Broadly speaking, the workflow provides a framework that yields
insight into the possible behavior of specific genes and the role they
play in functional pathways in response to treatment.  It takes
advantage of the current availability of high-throughput technologies,
which enable the access to transcriptomic data of organisms under
different conditions and a better understanding of their reaction
under different environmental stimuli.

The proposed approach is both a generalization and an extension of the
Weighted Gene Co-expression Network Analysis
(WGCNA)~\cite{langfelder2008wgcna,tian2018identifying}.  Like WGCNA,
the general idea behind the proposed approach is to identify, after a
sequence of normalization and filtering steps, specific modules of
overlapping communities underlying the co-expression network of genes.
The proposed approach is considered a \textit{generalization} of WGCNA
because module detection recognizes overlapping communities using the
Hierarchical Link Clustering (HLC)~\cite{ahn2010link} algorithm.
Conceptually, the generalization adds the overlapping nature of the
regulatory domains of the systems that generate the co-expression
network~\cite{gaiteri2014beyond}.  The intuition is that overlapping
modules allow for scenarios where biological components are involved
in multiple functions.
The workflow is also an \textit{extension} of WGCNA because two
additional constraints are considered: networks in the intermediate
steps are forced to be scale-free~\cite{barabasi2003scale} and LASSO
regression~\cite{tibshirani1996regression} selects the most relevant
modules of responsive genes.
The regularized regression technique of LASSO forces the coefficients
associated to the less relevant modules to be assigned the value
zero~\cite{desboulets2018review}; it is particularly useful in
scenarios where the number of variables is much larger than the number
of samples.
This condition is satisfied when the target variables represent the
overlapping communities (obtained with HLC) and the samples represent
genotype data, which is usually a small set due to the high cost of
the RNA sequencing process.  Finally, the proposed workflow is also
modular, since other module detection and selection techniques could
be explored instead of HLC and LASSO.

The approach was showcased with a systematic study on rice
(\textit{Oryza sativa}), a food source that is known to be highly
sensitive to salt stress~\cite{chang2019morphological}. RNA-seq data
was accessed from the GEO database~\cite{clough2016gene} (accession
number GSE98455). It represents $\numprint{57845}$ gene expression
profiles of shoot tissues measured under control and stress conditions
in $92$ accessions of the Rice Diversity Panel
1~\cite{eizenga2014registration}. A total of 6 modules were detected as
relevant in the response to salt stress in rice: 3 modules, each
grouping together 3 genes, are associated to shoot K content; 2
modules of 3 genes are associated to shoot biomass; and 1 module of 4
genes is associated to root biomass. These genes are potential targets
for experimental validation of salinity tolerance.  From the 19 genes,
16 are also identified as deferentially expressed for at least one of
the 92 accessions, which re-enforces the labeling of the genes as
stress responsive. Moreover, independent recent studies report that 5
of these 19 genes have been identified, through in vivo
experimentation, to saline stress. Other genes have GO-annotations
related to saline stress, or are reported to have conserved
heritability for both control and salt stress conditions.  Further
studies are needed to elucidate the detailed biological functions of
the remaining genes and their role in the mechanisms that respond to
salt conditions.

\paragraph{Paper Outline.}
The remainder of the paper is organized as follows. The
\hyperref[sec.prelim]{Preliminaries} section gathers foundations on
gene co-expression networks, HLC, and LASSO. The proposed workflow is
presented in \hyperref[sec.framework]{the Workflow} section, which
emphasizes on the logical steps of the data analysis process and the
internal structures supporting the approach. The
\hyperref[sec.case]{Case Study} section presents an application of the
workflow for the identification of rice genes that are sensitive to
salt stress. Finally, the \hyperref[sec.concl]{Concluding Remarks}
section draws some conclusions and future research directions.

\section*{Preliminaries}
\label{sec.prelim}

This section presents preliminaries on networks, the clustering
algorithm HLC, and the linear regression technique LASSO.

\subsection*{Co-expression network}

A \textit{network} is an undirected graph $G=(V,E)$ where
${V=\{v_1,v_2,\ldots,v_{n}\}}$ is a set of $n$ \textit{vertices} (or
\textit{nodes}) and ${E=\{e_1,e_2,\ldots,e_q\}}$ is a set of $q$
\textit{edges} (or \textit{links}) that connect vertices. In a
co-expression network of genes, each node corresponds to a gene and a
link indicates a common expression pattern between two genes.  The
network can be represented by an adjacency matrix $A \in \{0,1\}^{n
  \times n}$ that is symmetric. A matrix entry in positions
$(v_i,v_j)$ and $(v_j,v_i)$ is equal to $1$ whenever there is an edge
connecting vertices $v_i$ and $v_j$, and equal to $0$
otherwise. Co-expression networks are of biological interest because
adjacent nodes in the network represent co-expressed genes that are
usually controlled by the same transcriptional regulatory pathway,
functionally related, or members of the same pathway or metabolic
complex~\cite{FIONDA2019915}.

\subsection*{Hierarchical Link Clustering}

The Hierarchical Link Clustering (HLC) algorithm partitions groups of
links (rather than nodes), where each node inherits all memberships of
its links and can belong to multiple, overlapping
communities~\cite{ahn2010link}. More specifically, HLC evaluates the
similarity between links if they share a particular node.  Consider a
pair of incident links $e_{ik}$ and $e_{jk}$ to node $k$.  The
similarity between $e_{ik}$ and $e_{jk}$ is defined by the Jaccard
index as
\begin{equation}\label{eq:jaccard}
  S(e_{ik},e_{jk}) = \frac{\vert \ \eta(i) \cap \eta(j) \ \vert}{\vert \ \eta(i) \cup \eta(j) \ \vert},
\end{equation}
where $\eta(v)$ denotes the set containing the node $v$ and its
neighbors, for any $v \in V$. The algorithm uses single-linkage
hierarchical clustering to build a dendrogram where each leaf is a
link from the network and branches represent linked communities.

\begin{figure}[htbp]
  \centering
	  \captionsetup{type=figure}
    \includegraphics[clip,width=0.96\textwidth]{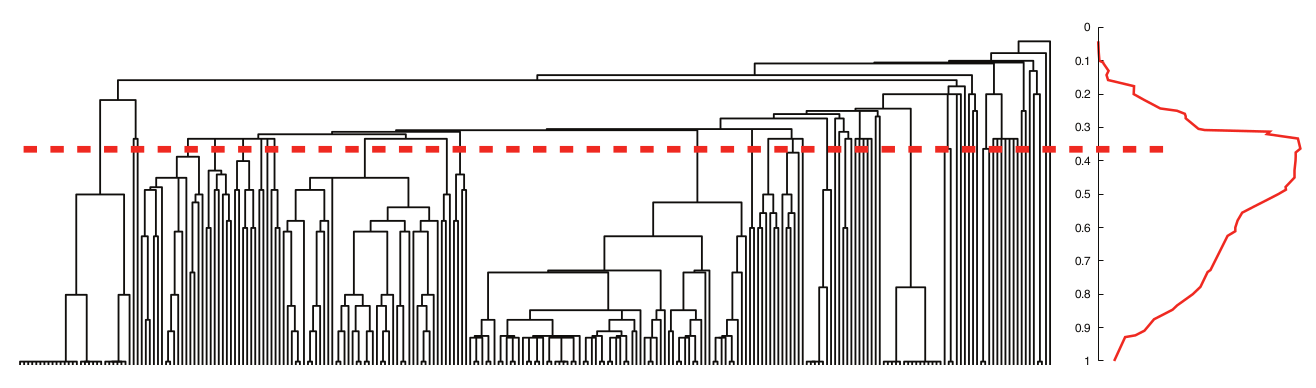}
  \caption[Example of a full link dendrogram (left) and partition density (right)]%
  {Example of a full link dendrogram (left) and partition density (right), borrowed from~\cite{ahn2010link}.}
  \label{fig:hlc_density}
\end{figure}

The threshold to cut the dendrogram is defined based on the average
density of links inside communities (i.e., partition density).  For
$G=(V,E)$ and a partition of the links into $c$ subsets, the partition
density is computed as
\begin{equation}
D = \frac{2}{\vert E \vert} \sum_c \vert E_c \vert \frac{\vert E_c \vert - \vert V_c \vert + 1 }{(\vert V_c \vert -1)(\vert V_c \vert -2)}.
\end{equation}
Note that, in most cases, the partition density $D$ has a single
global maximum along the dendrogram.  As depicted in
Figure~\ref{fig:hlc_density}, if the dendrogram is cut at the top,
then $D$ represents the average link density of a single giant
community. If the dendrogram is cut at the bottom, then most
communities consist of a single link. In other words, $D = 1$ when
every community is a clique and $D = 0$ when each community is a
tree. If a community is less dense than a tree (i.e., when the
community subgraph has disconnected components), then such a community
contributes negatively to $D$, which can take negative values. The
minimum density inside a community is $-2/3$, given by one community
of two disconnected edges. Since $D$ is the average of the
intra-community density, there is a lower bound of $-2/3$ for $D$. By
computing $D$ at each level of the dendrogram, the level that
maximizes partition density can be found (nonetheless, meaningful
structure could exist above or below the threshold).

The output of the cut is a set of node clusters, where each node can
participate in multiple communities.

\subsection*{Least Absolute Shrinkage Selector Operator (LASSO)}

LASSO is a regularized linear regression technique. By combining a
regression model with a procedure of contraction of some parameters
towards $0$, LASSO imposes a restriction (or a penalty) on regression
coefficients. In other words, LASSO solves the least squares problem
with restriction on the $ L_1$-norm of the coefficient vector. In
particular, the approach is especially useful in scenarios where the
number of variables $ c $ is much greater than the number of samples $
m $ (i.e., $ c \gg m $).

Consider a dataset of $m$ samples, consisting each of $c$ covariates
and a single outcome. Let $y_i$ be the outcome and $x_i :=
(x_{i1},...,x_{ic})$ be the covariate vector for the $i$-th
sample. The objective of LASSO is to solve
\begin{equation}
\min \left\lbrace\sum_{i=1}^{m}{\left( y_i-\sum_{j=1}^c{\alpha_j
    x_{ij}}\right)^2} \right\rbrace \quad , \quad \textrm{subject to}
\quad \sum_{j=1}^c \left\vert \alpha_j \right\vert \leq s,
\end{equation}
where $s$ is the regularization penalty.  Equivalently, in the
Lagrangian form, LASSO minimizes

\begin{equation}
\label{eq:lasso}
  \sum_{i=1}^{m}{\left( y_i-\sum_{j=1}^c{\alpha_j x_{ij}}\right)^2} +
  \lambda \sum_{j=1}^c \left\vert \alpha_j \right\vert ,
\end{equation}
where $\lambda \geq 0$ is the corresponding Lagrange multiplier. Since
the value of the regularization parameter $\lambda$ determines the
degree of penalty and the accuracy of the model, cross-validation is
used to select the regularization parameter that minimizes the
mean-squared error. LASSO is preferred in the proposed workflow
because it tends to outperform other methods such as ordinary least
squares regression and Ridge~{\cite{muthukrishnan2016lasso}}.

\section*{The Workflow}
\label{sec.framework}

\begin{figure}[htbp]
  \centering
    \captionsetup{type=figure}
    \includegraphics[clip,width=0.96\textwidth]{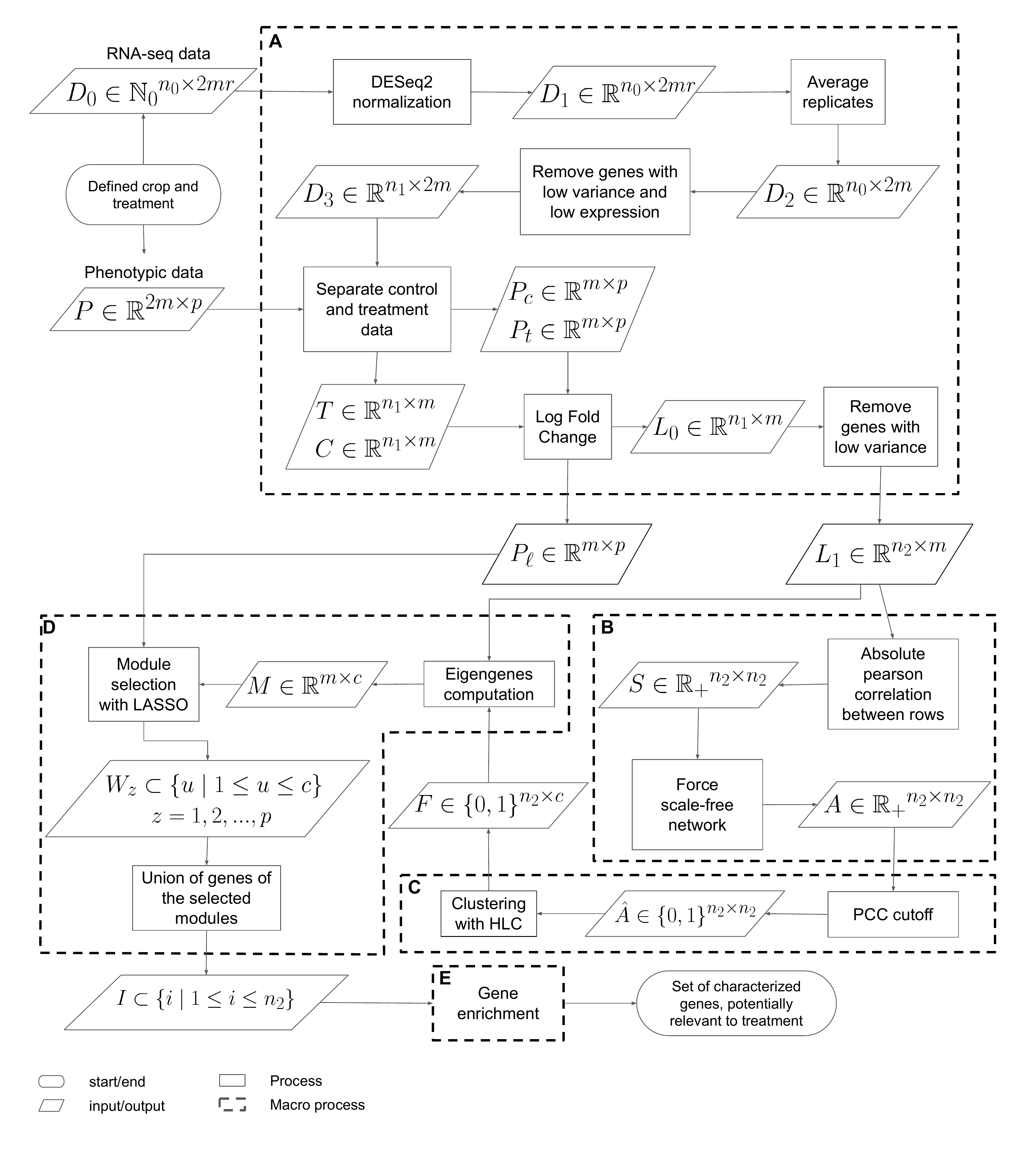}
  \caption[The proposed workflow is broken down into five macro-steps]%
  {The proposed workflow is broken down into five macro-steps:
    A.~Data pre-processing, B.~Co-expression network
    construction, C.~Identification of co-expression modules,
    D.~Detection of modules association to phenotypic traits, and
    E.~Gene enrichment.}
  \label{fig:flow_chart}
\end{figure}

Figure~\ref{fig:flow_chart} introduces the proposed workflow.  It can
be broken down into five macro-processes (A)-(E). Compared to WGCNA,
the workflow adds the macro-step (D) and generalizes macro-steps
(A)-(C).

The input of the workflow includes RNA-seq read counts, representing
gene expression levels. More precisely, the workflow uses $n_0$ gene
expression profiles measured for $m$ different genotypes of $r$
biological replicates (under control and treatment conditions). This
data is represented as matrix $D_0 \in {\mathbb{N}_0}^{n_0 \times
  2mr}$. To discover key genes and their interaction with phenotypes
related to treatment, the approach also requires a set of $p$
phenotypic traits measured for $m$ genotypes. The phenotypic data is
captured by matrix $P \in \mathbb{R}^{2m \times p}$, which contains
two phenotypic values per genotype (under control and treatment
conditions).

\subsection*{A. Data Pre-processing}

The goal of the data pre-processing stage is to build matrices
$P_\ell$ and $L_1$ representing, respectively, the changes in
phenotypic values and expression levels between control and treatment
condition. In other words, $P_\ell$ and $L_1$ are constructed from
RNA-seq and phenotypic data found in matrices $D_0$ and $P$.

A normalization process is applied to interpret RNA-seq data and
handle possible biases affecting the quantification of results. Here,
DESeq2~\cite{love2014moderated} is used to correct the library size
and RNA composition bias. The normalized data is represented as a
matrix $D_1 \in \mathbb{R}^{n_0 \times 2mr}$, and the biological
replicates of each genotype are averaged and represented as a matrix
$D_2 \in \mathbb{R}^{n_0 \times 2m}$. The genes exhibiting low
variance or low expression are removed from $D_2$. Consequently, this
stage of the approach reduces the set of genes from a pool of size
$n_0$ to a restricted pool of size $n_1 \leq n_0$.  The control and
treatment data is separated into the matrices $C\in \mathbb{R}^{n_1
  \times m}$ and $T\in \mathbb{R}^{n_1 \times m}$, respectively. The
matrix entries $c_{ij}$ in $C$ and $t_{ij}$ in $T$ represent the
normalized expression level of gene $i$ in accession $j$ under control
and treatment condition, respectively.  Control and treatment data is
also separated from phenotypic data $P$, obtaining two matrices $P_c$
and $P_t$ of dimensions $m \times p$.

In the above configuration, the changes in expression levels and
phenotypic values between control and treatment conditions are
measured in terms of logarithmic ratios. In the case of expression
levels, the log ratios are represented in the Log Fold Change matrix
$L_0 \in \mathbb{R}^{n_1 \times m}$, where $\ell_{ij}=\log_2
(t_{ij}/c_{ij})$. Similarly, the log ratios of the phenotypic data are
computed and represented in the $P_\ell \in \mathbb{R}^{m \times p}$
matrix.

The final stage of pre-processing is to filter $L_0$ by removing rows
(e.g., genes) with low variance in the differential expression
patterns, thus obtaining a new matrix $L_1$ of dimensions $n_2 \times
m$, with $n_2 \leq n_1$.

\subsection*{B. Construction of the Co-expression Network}

A gene co-expression network connects genes with similar expression
patterns across biological conditions. The purpose of this step is to
describe how to build the co-expression network $A$ from the Log Fold
Change matrix $L_1$: the goal is to capture the relationship between
genes according to the change in expression levels between the two
studied conditions. These co-expression patterns are meaningful for
the identification of genes that are not yet associated to treatment
response.

The Log Fold Change matrix $L_1$ is used to build the co-expression
network following the first two steps of
WGCNA~\cite{langfelder2008wgcna}.  First, the level of concordance
between gene differential expression profiles across samples is
measured. To this end, as proposed in WGCNA, the absolute value of the
Pearson Correlation Coefficient (PCC) is used as the similarity
measure between genes, meaning that pairs of nodes with strong
negative correlation are considered connected with the same strength
as nodes with strong positive correlation~{\cite{song2012comparison}}.
The resulting values are stored in the similarity matrix $S\in
\mathbb{R_{+}}^{n_2 \times n_2}$. Second, the matrix $S$ is
transformed into an adjacency matrix $A \in \mathbb{R_+}^{n_2\times
  n_2}$ where each entry $a_{ij} = (s_{ij})^\beta $ encodes the
connection strength between each pair of genes. In other words, the
elements of the adjacency matrix are the similarity values up to the
power $\beta > 1$ so that the degree distribution will fit a
scale-free network. These networks contain many nodes with very few
connections and a small number of hubs with high connections. In a
strict scale-free network, the logarithm of $P(k)$ (i.e., the
probability of a node having degree $k$) is approximately inversely
proportional to the logarithm of $k$ (i.e., the degree of a node).
The parameter $\beta$ is chosen to be the smallest value for which the
$R^2$ of the linear regression between $log_{10}(p(k))$ and
$log_{10}(k)$ is closest to $1$ (here, $R^2 > 0.8$).

\subsection*{C. Identification of Co-expression Modules}

The next step in the workflow is to identify modules of overlapping
communities from the co-expression network represented by $A$.  The
idea is to cluster genes with similar patterns of differential
expression change. Membership in these modules may overlap in
biological contexts, because modules may be related to specific
molecular, cellular, or tissue functions, and the biological
components (i.e., genes) may be involved in multiple functions. Unlike
WGCNA, the workflow applies the Hierarchical Link Clustering (HLC)
algorithm (overviewed in the \hyperref[sec.prelim]{Preliminaries}
section) to detect overlapping rather than non-overlapping
communities.

First, the adjacency matrix $A$ is transformed into an unweighted
network $\hat{A} \in \{0,1\}^{n_2 \times n_2}$.  To this end, the PCC
cutoff is determined using the approach described
in~\cite{aoki2007approaches}. The number of nodes, edges, and the
network density is determined for different PCC cutoffs.  In a
neighborhood of the optimal PCC cutoff, the number of nodes presents a
linear decrease and the density of the network reaches its minimum,
while below this value the number of edges rapidly
increases. Following this observation, a cutoff is selected such that
gene pairs having a correlation score higher than the threshold are
considered to have a significant level of co-expression. The entries
of $A$ become $1$ above the cutoff and $0$ otherwise. The HLC
algorithm organizes the $n_2$ genes of matrix $\hat{A}$ into
$c$~modules, where each gene can belong to zero or multiple modules.
This information is represented as an affiliation matrix $F \in
\{0,1\}^{n_2 \times c}$, where $f_{iu} = 1$ if node $i$ is a member of
module $u$ (and $f_{iu}=0$, otherwise).

\subsection*{D. Detection of Modules Association to Phenotypic Traits}

Each module is represented by an eigengene, which is defined as the
first principal component of such module. An eigengene can be seen as
an average differential expression profile for each community: it is
computed from the Log Fold Change Matrix $L_1$ and the affiliation
matrix $F$. Given a module $u$, the affiliation matrix $F$ is used to
identify the genes belonging to $u$.  The corresponding rows of the
matrix $L_1$ are selected to compute the first principal component of
$u$. Each principal component becomes a column of the matrix $M \in
\mathbb{R}^{m \times c}$.

These profiles are associated with each phenotypic trait using LASSO
as a feature selection mechanism~{\cite{fonti2017feature}}.
Therefore, to identify the most relevant modules associated with the
phenotypic response to the specific treatment, the eigengenes (i.e.,
the columns of $M$) act as regressor variables and each phenotypic
trait (i.e., each column of $P_\ell$) is used as an outcome variable.
LASSO is applied $z \in \{1,2,...,p\}$ times, once for each phenotypic
trait. Recall that $y_i$ in Equation~{\ref{eq:lasso}} is the
phenotypic response for the $i$-th sample $(i \in \{1,2,...,m\})$,
$x_{ij}$ is the $i$-th value of the eigengene that represents the
$j$-th module $(j \in \{1,2,...,c\})$, and the weight $\alpha_j$
represents the importance of the $j$-th module in the phenotypic
response. The regularization parameter $\lambda$, tuned with
cross-validation, determines the number of modules to be selected.
The weights $\alpha$ evolve with each LASSO iteration, by trying to
minimize the value of Equation~{\ref{eq:lasso}}, until the desired
number of modules with non-zero weight is found. Intuitively, the
repetitive use of LASSO in the workflow achieves the goal of
neglecting (i.e., reducing to zero) the weights associated to modules
with non-essential effects in the phenotypic response and, at the same
time, enhancing the weights associated to modules with significant
effects.

The output after the repetitive application of LASSO is a set $W_z$ of
modules for each phenotypic trait $z$, where $W_z \subseteq \{u \mid 1
\leq u \leq c\}$ for $z= 1,2,..,p$. A target gene in $I$ for
downstream analysis is any gene belonging to a selected module; that
is, $I = \cup_{z=1}^{p} W_z$, where $I \subseteq \{i \mid 1 \leq i
\leq n_2\}$.

\subsection*{E. Gene Enrichment}

This is the final step of the workflow. Its goal is to annotate with
additional information the genes identified in previous stages,
helping to elucidate their possible behavior and role in the response
to the treatment under study.

A crucial step is to identify the differentially expressed genes in
the set $I$. That is, to select the genes in $I$ that have an absolute
value of the Log Fold Change of at least $2$ ($|\ell_{ij}|\geq 2$) for
at least one sample. This corresponds to genes whose expression level
is quadrupled (up or down) from the control to treatment condition;
they are the target genes.

Furthermore, functional category enrichment can be carried out by,
e.g., searching for gene ontology annotations in databases such as
QuickGO~\cite{binns2009quickgo}, UniProt~{\cite{uniprot2019uniprot}},
and the Rice Genome Annotation
Project~{\cite{kawahara2013improvement}}.  Such annotations can
provide evidence of biological implications of the target genes in the
treatment-tolerance mechanisms. Furthermore, those databases can be
used to identify the protein products of genes, which can be used in
turn to provide new insights on how target genes are involved in
functional pathways related to treatment.  Such analysis includes a
review of reported protein-protein interactions in databases such as
STRING~\cite{szklarczyk2016string}. The protein interactions include
direct (physical) and indirect (functional) associations. They stem
from computational prediction, knowledge transfer between organisms,
and interactions aggregated from other (primary) databases. The search
for unknown interactions would extend the workflow with additional
steps.

\section*{Identifying Potential Saline Stress Responsive Genes in Rice}
\label{sec.case}

This section presents a case study, applying the approach introduced
in \hyperref[sec.framework]{the Workflow} section, for identifying
genes that respond to saline stress in \textit{Oryza sativa}.  The
goal of this case study is to discover groups of genes whose
differential expression patterns are highly related to phenotypic
responses to salt stress. The discovery process is validated with a
Fisher's exact test, thus ensuring that the number of differentially
expressed genes (DEG) and of reported genes related to salt stress is
statistically significant.

The RNA-seq data was accessed from the GEO database
\cite{clough2016gene} (accession number GSE98455). It corresponds to
$n_0=\numprint{57845}$ gene expression profiles of shoot tissues
measured for control and salt conditions in $m=92$ accessions of the
Rice Diversity Panel 1~\cite{eizenga2014registration}, with $r=2$
biological replicates. A total of $p=3 $ phenotypic traits were used:
shoot $K$ content, and root and shoot biomass. These traits were
measured for the same $92$ genotypes, under control and treatment
conditions, and can be found in the supplementary information
for~\cite{campbell2017allelic}.

\subsection*{A. Data Pre-processing}

DESeq2 normalization was applied to the raw data and the biological
replicates were averaged. Genes exhibiting low variance were
identified as those with ratio of upper quantile to lower quantile
smaller than $1.5$ and were removed from the normalized data. Genes
with low expression, corresponding to those having more than $80\%$
samples with values smaller than $10$, were also removed. After this
filtering process a total of $n_1 = \numprint{9414}$ genes were kept
for further analysis.

Genes whose difference between the upper and lower quantiles was
greater than $0.25$ were removed from the Log Fold Change matrix
$L_0$. Therefore, the resulting matrix $L_1$ contained the log ratios
of $n_2 = \numprint{8928}$ genes. The logarithmic ratios of the
phenotypic data, for the $92$ accessions and the $3$ traits, were also
computed.

\subsection*{B. Construction of the Co-expression Network}

\begin{figure}[htbp]
  \centering
  	 \captionsetup{type=figure}
    \includegraphics[clip,width=0.8\textwidth]{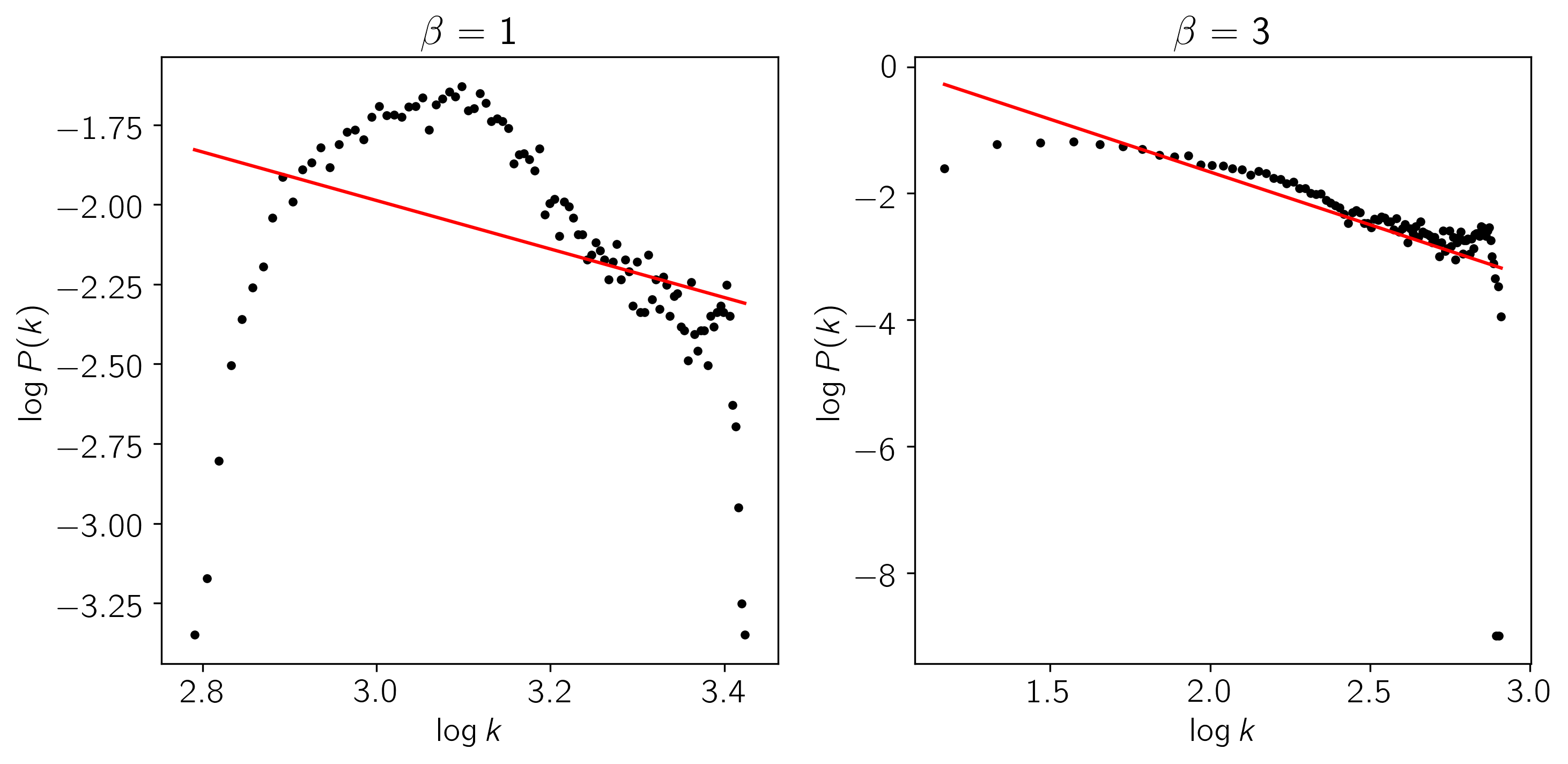}
  \caption{Degree distribution of the network represented by $S$ (left) and $A$ (right).}
  \label{fig:beta}
\end{figure}

The Log Fold Change matrix $L_1$ was used to compute the corresponding
similarity matrix.  For this network, it was observed that $\beta=3$
is the smallest integer such that $R^2 \geq 0.8$.
Figure~\ref{fig:beta} depicts the degree distribution of the
similarity matrix (left) and the degree distribution of the adjacency
matrix (right), which is the degree distribution of a scale-free
network with $R^2 = 0.8$ and $\beta = 3$.

The resulting adjacency matrix $A$ represents a complete graph
$G=(V,E)$, with $|V| = \numprint{8928}$ genes ($|E| =
\numprint{39850128}$ edges).

\subsection*{C. Identification of Co-expression Modules}

The adjacency matrix $A$ was transformed into an unweighted network
$\hat{A}$ applying the approach described
in~\cite{aoki2007approaches}. The cutoff value was set to $0.2$, based
on the density of the network combined with the decreasing number of
nodes and edges with higher PCC values. Hence, only connections above
this threshold were kept, while isolated nodes were removed. The
resulting adjacency matrix $\hat{A}$ consists of $\numprint{5810}$
connected genes and accounts for $ \numprint{614501}$ edges.

\begin{figure}[htbp]
  \centering
     \captionsetup{type=figure}
    \includegraphics[clip,width=0.95\textwidth]{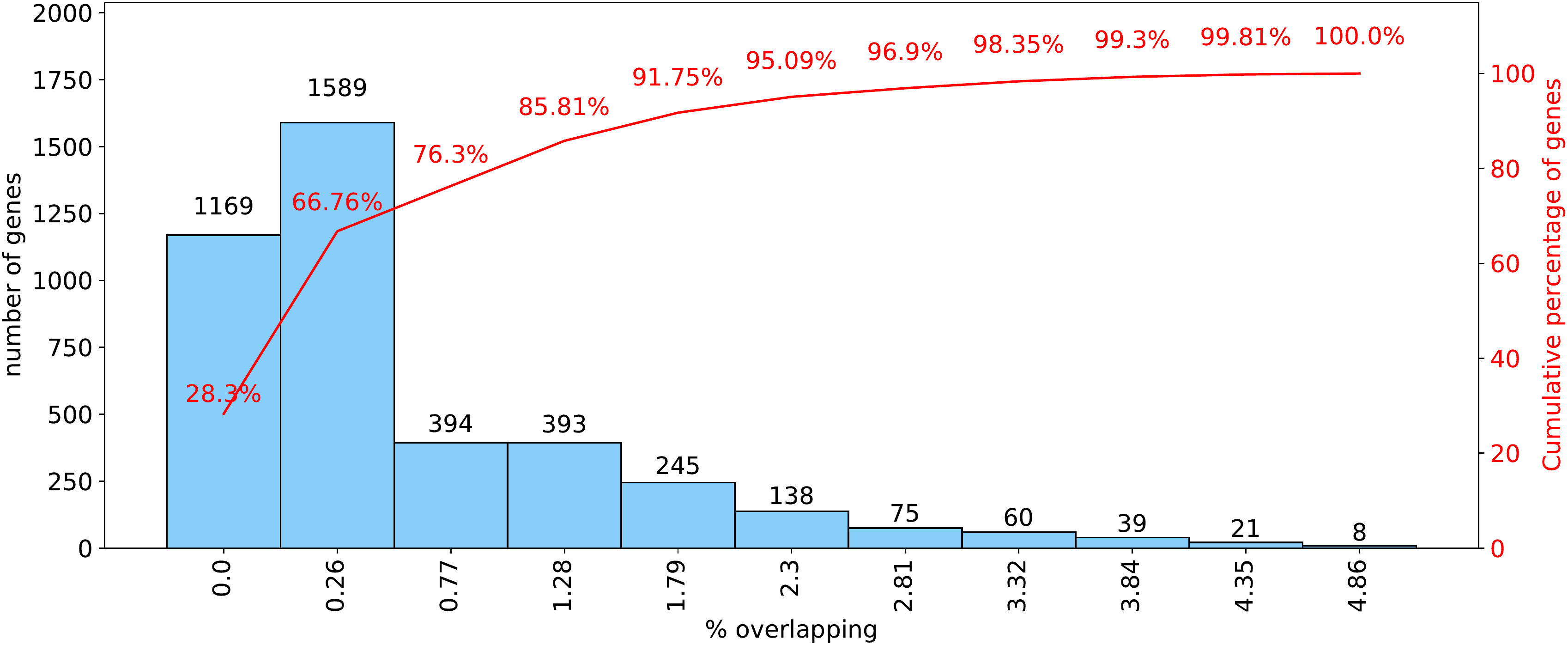}
  \caption{Overlapping percentage of genes after applying HLC.}
  \label{fig:overlap}
\end{figure}

The HLC algorithm distributes $\numprint{4131}$ genes in $c =
\numprint{5143}$ overlapping modules of at least $3$ genes
each. Figure~\ref{fig:overlap} presents a histogram of the overlapping
percentage of these genes, measured as the proportion of modules to
which each gene belongs. The first bar of the histogram represents the
genes with zero overlap, corresponding to $28\%$ of the total genes;
the remaining $72\%$ genes belong to more than one module.

\subsection*{D. Detection of Module Association to Phenotypic Traits}

\begin{figure}[htbp]
  \centering
     \captionsetup{type=figure}
    \includegraphics[clip,width=0.9\textwidth]{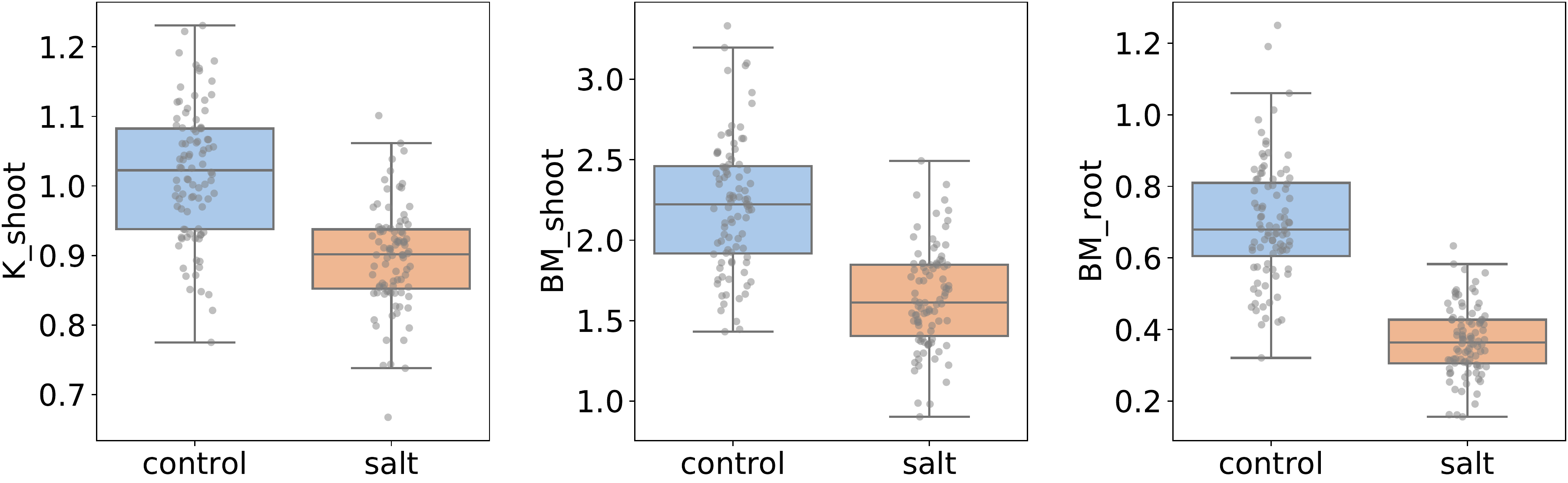}
  \caption{Phenotypic traits distribution under control and salt stress.}
  \label{fig:pdata}
\end{figure}

The phenotypic traits under study are shoot $K$ content, and root and
shoot biomass. Figure~\ref{fig:pdata} suggests that there are
significant differences in the values of these phenotypic traits
between stress and control conditions. This supports the working
hypothesis that these three variables represent tolerance-associated
traits in rice under salt stress.

\begin{figure}[htbp]
  \centering
  	 \captionsetup{type=figure}
    \includegraphics[clip,width=0.96\textwidth]{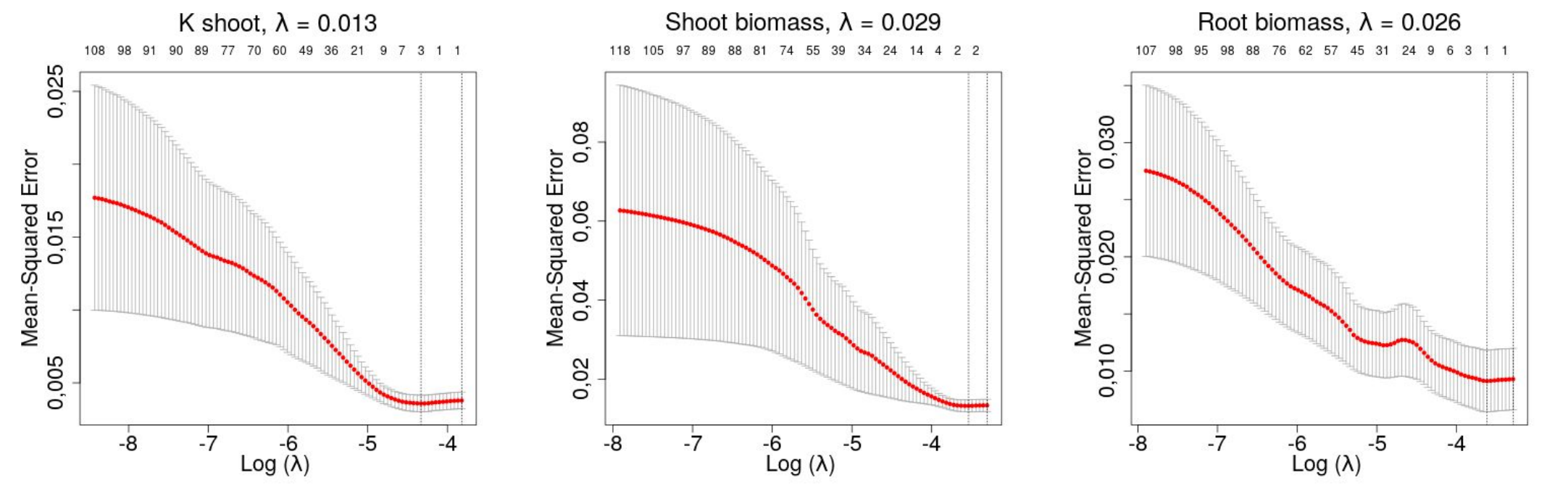}
  \caption{Cross-validation of the LASSO regularization parameter
    $\lambda$, for each phenotypic trait.}
  \label{fig:cross-val}
\end{figure}

By using the affiliation matrix $F$ derived from the HLC output and
the Log Fold Change matrix $L_1$, a matrix $M$ was built by computing
the eigengene for each of the $c = \numprint{5143}$ modules. LASSO was
applied by using each of the phenotypic traits as the outcome
variable, one at a time. As shown in Figure~\ref{fig:cross-val},
cross-validation was performed for each phenotypical trait to select
the corresponding regularization parameter $\lambda$ minimizing the
mean-squared error.

\begin{figure}[htbp]
  \centering
  	 \captionsetup{type=figure}
    \includegraphics[clip,width=0.8\textwidth]{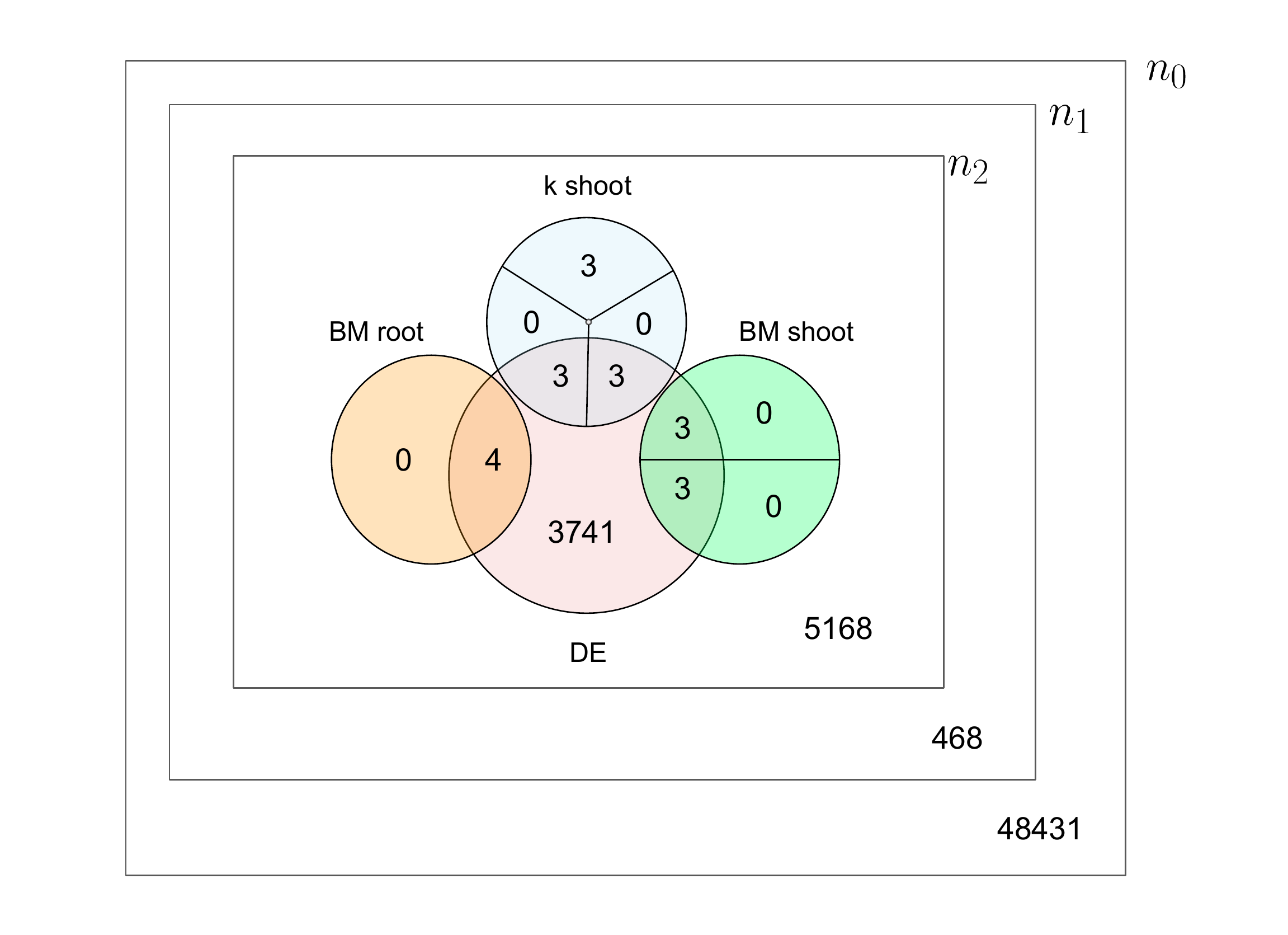}
   \caption[Venn diagram for the case study in rice]%
   {Venn diagram representing the number of genes selected at
    different stages of the proposed workflow for the case study in
    rice.}
  \label{fig:final_genes}
\end{figure}

Three LASSO models were adjusted by using the corresponding $\lambda$
and phenotypical data with the eigengenes of matrix $M$. As result, 6
modules were detected as relevant in the response to salt stress in
rice: 3 modules of 3 genes, each associated with shoot $K$ content; 2
modules of 3 genes associated with shoot biomass; and 1 module of 4
genes associated with root biomass. Figure~\ref{fig:final_genes}
depicts in a Venn diagram how the number of genes selected at
different stages evolved.

\subsection*{E. Gene Enrichment}
\label{subsec:enrich}

From the $19$ genes selected by LASSO, $16$ genes ($84\%$) were also
identified as differentially expressed ($|\ell_{ij}| \geq 2$) for at
least one of the $92$ accessions.  In general, there were
$\numprint{3741}$ unselected differentially expressed genes and
$\numprint{5168}$ unselected non-differentially expressed ones, for a
total of $\numprint{8909}$ genes. Therefore, differentially expressed
genes were significantly more likely to be selected by the workflow,
as checked by a Fisher exact test with p-value less than $10^{-3}$.

Figure~\ref{fig:final_genes} summarizes how, from the initial
$n_0=\numprint{57845}$ genes, the proposed workflow identified a
reduced set of $19$ genes. First, $\numprint{48431}$ genes were
discarded after filtering the normalized expression data $D_2$ and
then $486$ additional genes were discarded when filtering the Log Fold
Change matrix $L_0$. A final set of $19$ genes are identified, of
which $16$ are differentially expressed.

The 19 selected genes were also enriched by contrasting them with
findings reported in the 
literature~{\cite{du2019network,lahiri2021bayesian,mcgowan2021chromosomal,yu2021genome}}, 
which applied different approaches to study the same RNA-seq dataset
GSE98455.  In~{\cite{mcgowan2021chromosomal}}, 11 of the 19 selected
genes were reported to have conserved heritability for both control
and salt stress conditions.

The identifiers for the 19 genes are listed in
Table~\ref{tab:genes}. Differentially expressed genes are identified
by the mark (*) in column DEG, and those with heritable expression
under control and salt stress (as reported in the literature) in
column H.

\begin{table}[htbp]
\centering
\caption{Selected genes}
\label{tab:genes}
\begin{tabular}{cccccc}
\hline
Phenotypic trait           & Module              & TU ID        & LOC\_Os ID      & DEG & H \\ \hline \hline
\multirow{9}{*}{K\_shoot}  & \multirow{3}{*}{1}  & 13101.t01457 & LOC\_Os01g16124 & *   & * \\
                           &                     & 13101.t01458 & LOC\_Os01g16130 & *   & * \\
                           &                     & 13104.t01366 & LOC\_Os04g16230 & *   &   \\ \cline{2-6} 
                           & \multirow{3}{*}{2}  & 13104.t01068 & LOC\_Os04g12520 & *   & * \\
                           &                     & 13104.t01069 & LOC\_Os04g12530 & *   & * \\
                           &                     & 13104.t01066 & LOC\_Os04g12499 & *   & * \\ \cline{2-6} 
                           & \multirow{3}{*}{3}  & 13101.t00913 & LOC\_Os01g10400 &     &   \\
                           &                     & 13102.t03795 & LOC\_Os02g41820 &     & * \\
                           &                     & 13103.t00468 & LOC\_Os03g05870 &     & * \\ \hline
\multirow{4}{*}{BM\_shoot} & \multirow{4}{*}{4}  & 13101.t02836 & LOC\_Os01g33450 & *   & * \\
                           &                     & 13102.t01261 & LOC\_Os02g14520 & *   &   \\
                           &                     & 13107.t03589 & LOC\_Os07g39390 & *   &   \\
                           &                     & 13112.t00905 & LOC\_Os12g10280 & *   &   \\ \hline
\multirow{6}{*}{BM\_root}  & \multirow{3}{*}{5}  & 13101.t05133 & LOC\_Os01g58100 & *   &   \\
                           &                     & 13112.t02444 & LOC\_Os12g27254 & *   &   \\
                           &                     & 13112.t03421 & LOC\_Os12g37260 & *   & * \\ \cline{2-6} 
                           & \multirow{3}{*}{6}  & 13104.t03155 & LOC\_Os04g35010 & *   & * \\
                           &                     & 13108.t03971 & LOC\_Os08g42310 & *   & * \\
                           &                     & 13109.t01501 & LOC\_Os09g17049 & *   &   \\ \hline
\end{tabular}
\centering
\end{table}

Salinity tolerance comes from genes that limit the rate of salt uptake
from the soil and the transport of salt throughout the plant, adjust
the ionic and osmotic balance of cells in roots and shoots, and
regulate leaf development and the onset of
senescence~{\cite{munns2005genes}}.
GO terms related to these characteristics, and therefore relevant to
salt stress, were found in this case study to be associated with some
selected genes. For example, gene LOC\_Os12g37260 is annotated with
response to abiotic stimulus and response to stress, and gene
LOC\_Os12g10280 is annotated with response to extracellular stimulus,
channel activity, and transmembrane transport.  Genes LOC\_Os04g12499,
LOC\_Os04g12530, and LOC\_Os12g10280 are annotated with transporter
activity, while gene LOC\_Os04g35010 is annotated with multicellular
organism development.

\begin{figure}[htbp]
  \centering
  	 \captionsetup{type=figure}
    \includegraphics[clip,width=0.8\textwidth]{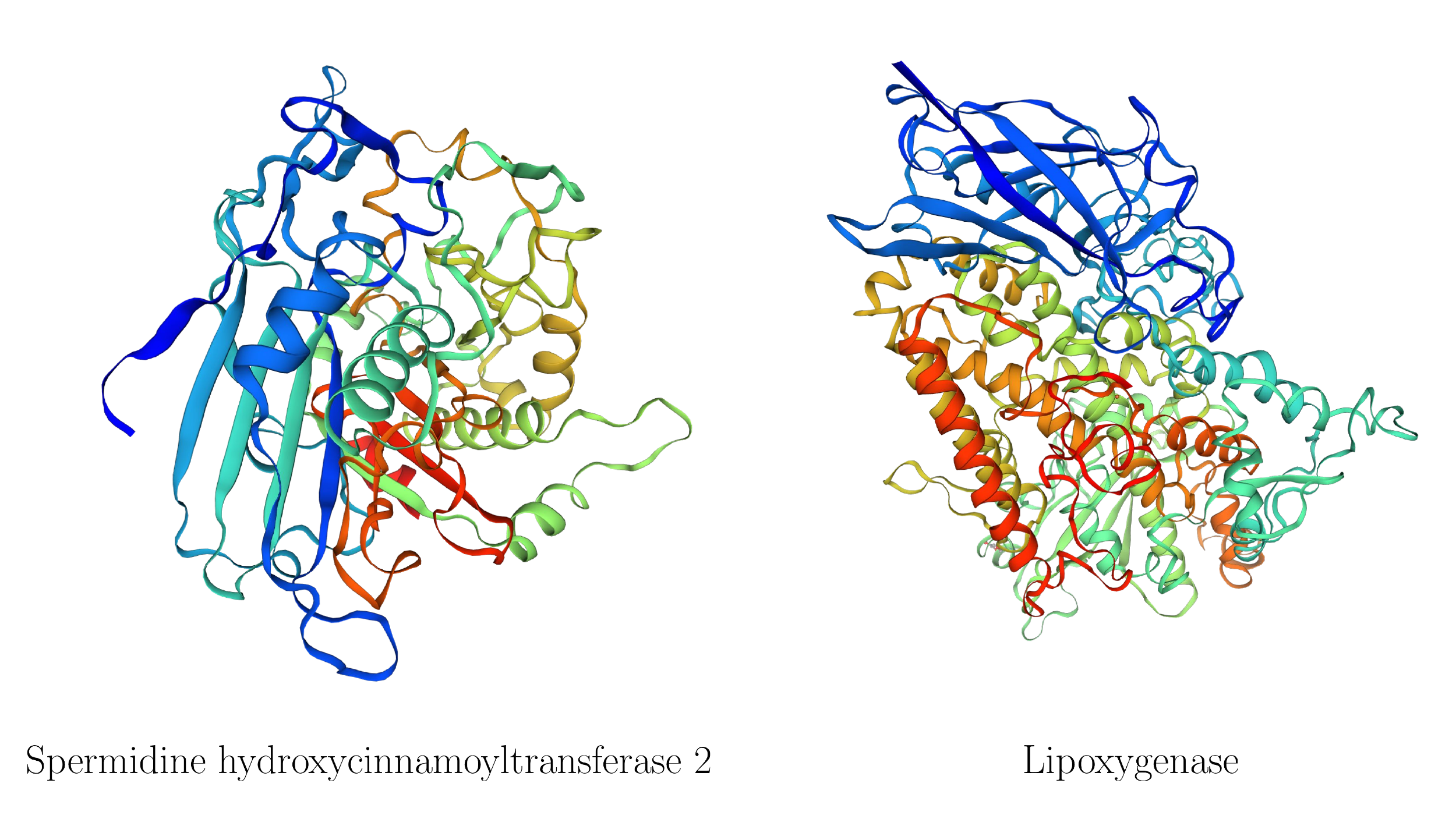}
  \caption{3D protein structure of named genes selected by LASSO, borrowed from~\cite{szklarczyk2016string}.}
  \label{fig:3d}
\end{figure}

In vivo experiments, reported by independent authors, provide evidence
on the relationship with salt stress of 5 genes among the ones
selected in the case study ($26\%$).
Gene LOC\_Os04g12530 is reported as an up-regulated gene in rice
plants tolerant to salt stress~{\cite{razzaque2019gene}}.
Gene LOC\_Os12g10280 encodes an aquaporin nodulin 26-like intrinsic
membrane (NIP3;5) protein~{\cite{hsieh2018early}}; it has been shown
that NIPs play an important role in salt stress responses and in
maintaining plant water balance~{\cite{kapilan2018regulation}}.
Gene LOC\_Os04g35010 encodes a protein from the bHLH domain, which
have been shown to be part of multiple cellular processes, including
salt stress signaling pathways~{\cite{qian2021regulatory}}.
Gene LOC\_Os12g27254 encodes spermidine hydroxycinnamoyltransferase 2
(SHT2) protein. This protein contributes to the natural variation of
spermidine-based phenolamides in rice cultivars, which is known to
promote tolerance to saline
stress~{\cite{bassard2010phenolamides,roychoudhury2011amelioration,gupta2013plant,peng2019novel}}.
Gene LOC\_Os12g37260 encodes the Lipoxygenase protein, which is known
to correlate directly with salt tolerance in
rice~{\cite{mittova2002salt,mostofa2015trehalose,hou2015persimmon}}.
Note that the STRING database reported a protein-protein interaction of
the last two mentioned proteins, namely SHT2 and Lipoxygenase, supporting
their membership within the same module, as seen in
Table~{\ref{tab:genes}}. Figure~\ref{fig:3d} shows the
corresponding 3D protein structures of these two proteins.
In relation to the $5$ genes above-mentioned, there are $387$ other
genes known to be involved in salt stress (see
~{\cite{razzaque2019gene,liu2019identification,chen2020genome}}).
Therefore, it can be said that the number of genes selected by the
workflow that are related to salt stress is significant, as checked by
a Fisher exact test with p-value less than $10^{-2}$.

As a conclusion, the results presented in this section strongly
suggest that the proposed workflow, based on identifying overlapping
communities in co-expression networks, is capable of detecting stress
responsive genes.  Further studies are needed to elucidate the
detailed biological function of the remaining 14 genes --out of the
initial $\numprint{57845}$ genes-- that have not been reported in the
literature to be related to salt stress response.  This study suggests
that they may have the potential to intervene in stress responsive
mechanisms to salt conditions in rice.

\section*{Concluding Remarks}
\label{sec.concl}

This manuscript provides a detailed description of a network-based
analysis workflow for the discovery of key genes responding to a
specific treatment in plants. It links transcriptomic with
phenotypic data and identifies overlapping gene modules.

The proposed approach was inspired by the workflow suggested in the
WGCNA~\cite{langfelder2008wgcna}. Its main steps are the preprocessing
of the gene expression data, the construction of a co-expression
network, the detection of modules within the network, the relation of
modules with external information (e.g., phenotypic data), and the
enrichment of the identified key genes with additional information.
Both approaches are structured in a modular way, which allows
modifying and exploring different techniques in each step of the
workflow.

The proposed workflow is designed to integrate expression data
measured under two different conditions (namely, control and
treatment), unlike the usually co-expression-based approaches working
with both conditions independently or considering only a single
condition. For this purpose, an approach similar to that proposed
in~\cite{du2019network} is used, where the control and treatment data
are compiled in a single matrix using the Log Fold Change
measure. Thus, the input to construct the co-expression network is not
the expression data, but instead the changes in the expression levels
from one condition to the other, making room for capturing the signal
of changes caused by the treatment.

An important feature in the proposed workflow is the module detection
technique. The co-expression network is computed, as in WGCNA, until a
scale-free network is obtained. In the proposed approach, this network
is then used to apply the HLC algorithm, a clustering tool capable of
detecting overlapping communities. Several approaches of module
detection from gene expression have been proposed and are evaluated
in~\cite{saelens2018comprehensive}. Most of them focus mainly on
disjoint (non-overlapping) communities; the techniques described to
deal with overlaps are not clustering, but bi-clustering and
decomposition methods. It is well known that communities in real
networks, including biological ones, are likely to
overlap~\cite{palla2005uncovering}. Thus, the approach presented in
this work can be seen as a generalization of the previous approaches,
such as WGCNA, with the potential to deal with genes associated to
multiple biological processes.

The workflow proposed in this paper was applied in a case study with
rice under salt stress. It identified a group of $19$ genes, of which
$16$ were differentially expressed and $5$ have been reported to be
related to saline stress response in independent in vivo experiments
by other
authors~\cite{razzaque2019gene,kapilan2018regulation,qian2021regulatory,bassard2010phenolamides,roychoudhury2011amelioration,mostofa2015trehalose}.  Moreover, also
$5$ of the $19$ genes have GO-annotations related to saline stress,
and $11$ genes are reported to have conserved heritability for both
control and salt stress conditions.

As future work, other overlapping module detection and selection
techniques should be used, complementing HLC and LASSO,
respectively. The combination of these techniques would allow finding
target genes for future biological studies that evaluate their
potential as genes that respond to salt stress in rice, and other
crops and stresses. In vivo laboratory experimentation needs to be
conducted to validate the findings of this paper in relation to
salinity stress for some of the $19$ genes.

Finally, the workflow is presented as a protocol capable of
considerably reducing the number of genes detected as relevant in the
response to a given stress. Other traditionally used methods for
this purpose tend to generate a large list of candidate genes, thus
limiting subsequent efforts in experimental validation. In this sense,
the proposed workflow can help in reducing such efforts in time and
money invested by researchers in the experimental validation of
stress-responsive genes.

\section*{Abbreviations}
\begin{description}
\item[GO:] Gene Ontology 
\item[HLC:] Hierarchical Link Clustering 
\item[LASSO:] Least Absolute Shrinkage Selector Operator 
\item[PCC:] Pearson Correlation Coefficient 
\item[RNA-seq:] RNA sequencing 
\item[SHT2:] Spermidine hydroxycinnamoyltransferase 2 
\item[WGCNA:] Weighted Gene Co-expression Network Analysis
\end{description}

\section*{Availability of data and materials}
The datasets analyzed for the current study are publicly available from
different sources. They can be found in the following locations:

\begin{itemize}

\item RNA-seq data of salt stress in rice is available on the GEO
  (GSE98455).

\item Phenotypic data of salt stress in rice is a subset of the
  supplementary file 1 included in~\cite{campbell2017allelic}.
\end{itemize}

The data collected, cleaned, and processed from the above sources as
used in the case study can be requested to the authors.

A workflow implementation is publicly available:
\begin{itemize}
\item
  Project name: Condition-specific co-expression network analysis
\item
  Project home page: \url{https://github.com/criccio35/workflow_stress}
\item
  Operating system(s): platform independent.
\item
  Programming language: Python 3.
\item
  Other requirements: None.
\item
  License: GNU GPL v3.
\end{itemize}

\bibliographystyle{splncs04}
\bibliography{biblio}

\end{document}